\documentclass[twocolumn,reprint,amsmath,amssymb]{revtex4}

\usepackage{graphicx}
\usepackage{dcolumn}
\usepackage{bm}

\def\BEq{\begin{equation}}
\def\EEq{\end{equation}}
\def\BEqA{\begin{eqnarray}}
\def\EEqA{\end{eqnarray}}
\def\BEn{\begin{enumerate}}
\def\EEn{\end{enumerate}}
\def\BWT{\begin{widetext}}
\def\EWT{\end{widetext}}
\def\a{\alpha}

\def\d{\delta}

\def\m{\mu}

\def\n{\nu}

\def\r{\rho}
\def\s{\sigma}

\def\w{\omega}

\allowdisplaybreaks[1]

\begin{document}


\title{Neutron interference in the Earth's gravitational field}

\author{Andrei Galiautdinov$^1$ and Lewis H.\ Ryder$^2$}
 \affiliation{
$^1$Department of Physics and Astronomy, 
University of Georgia, Athens, GA 30602, USA\\
$^2$School of Physical Sciences, University of Kent, Canterbury, Kent, CT2 7NH,
United Kingdom}

\date{\today}

\begin{abstract}

This work relates to the famous experiments, performed in 1975 and 1979 by Werner et al., measuring neutron interference and neutron Sagnac effects in the earth's gravitational field. Employing the method of Stodolsky in its weak field approximation, explicit expressions are derived for the two phase shifts, which turn out to be in agreement with the experiments and with the previously obtained expressions derived from semi-classical arguments: these expressions are simply modified by relativistic correction factors.

\end{abstract}


\maketitle


\allowdisplaybreaks[1]

\section{Introduction}

It is now several decades since the ground-breaking work by Werner and his
co-workers showed that gravitational \cite{1.0, 1} and rotational \cite{2} effects were to be
found in neutron interference experiments performed on the earth's surface \cite{2.1, 2.2, 2.3}. The
predicted and experimentally confirmed gravitational phase shift is the only
expression in physics to feature both Newton's constant of gravitation $G$ and
Planck's quantum of action $\hbar$, which surely makes these experiments particularly
noteworthy. The two experiments are referred to hereafter as the COW experiment and
the neutron Sagnac effect.

Straightforward, semi-classical derivations of these effects have already appeared
in the literature (see for example \cite{3, 3.1, 4}) and in abbreviated form are summarized 
in Sections \ref{sec:SimpleCOW} and \ref{sec:SimpleSAGNAC} below. What is
very clear, however, is that a proper account of this topic should really be sought
in General Relativity (GR) --- which is, after all, a theory of gravity! --- and indeed
numerous papers have been written using this approach (see for example \cite{5,6,6.1,7,8}). 
Some of these explore rather sophisticated notions, for example a possible parallel between the COW experiment and the Aharonov-Bohm effect, based on the integrated curvature of an enclosed path, on the one hand in parameter space and on the other hand in field space \cite{7}. We do not aim to explore these higher-flown topics, but rather
to present a simple demonstration of how GR can account for the findings in neutron
interferometry, and, at an introductory level suitable for instance for inclusion
in an introductory course, demonstrate that general relativity has an application
in quantum physics \cite{SAKURAI93} --- a notion which might still cause some surprise!

We use the Kerr solution of GR \cite{KERR63, BL67}, 
since this includes the rotation of the earth through
the angular momentum parameter $a$, as well as $\omega$, the angular velocity of the earth, and $r_s$ its
Schwarzschild radius. These are all small parameters, and we calculate
the relevant effects to second order in all these quantities (mixed and unmixed).
The general method of procedure is the weak field approximation, adopted by
Stodolsky \cite{5}.

The next Section describes a standard, elementary derivation of the COW
effect, and in Section \ref{sec:SimpleSAGNAC} is a similarly elementary 
derivation of the Sagnac effect
for neutrons. In Section \ref{sec:Kerr_metric} 
the general relativistic setting for more realistic
derivations of these effects is presented. The Kerr metric is displayed as well as a
coordinate transformation to a Cartesian system relevant to our problem. In the
final Section our results are derived. Use is made of the weak field approximation
in conjunction with a specific assumption which allows the calculations to be
preformed. It is found that the resulting phase shifts are, in both cases, those
predicted by the simple models in Sections \ref{sec:SimpleCOW} 
and \ref{sec:SimpleSAGNAC}, with correction factors of
$\gamma = (1- {v^2}/{c^2})^{-{1/2}}$, and additional small terms involving 
$\omega$, the angular velocity of the earth.

\section{Simple derivation of COW effect}
\label{sec:SimpleCOW}

The setup described in reference \cite{1} (see also \cite{4}) is based on the splitting of the
neutron beam by Bragg diffraction from perfect crystals, as first implemented for X
rays by Bonse and Hart \cite{6new}. Rauch and Werner \cite{4} point out that when the desired degree of crystal cutting is achieved, the resulting interferometry "exhibits the fundamentals of quantum mechanics in a very direct and obvious way". The interference involved is "topologically equivalent to a ring", which we represent as a rectangle, of macroscopic dimensions (centimetres). The neutrons enter at the bottom left corner where the beam splits
into two, and the beams recombine at the top right corner, where the interference
takes place.

The spatial part of a plane matter wave describing a neutron beam is given by 
$e^{i\bold{k \cdot {r}}}$, where $\bold{k}$ is the wave vector and $k \equiv |\bold{k}| = 2\pi/\lambda$ is the wave number, 
with $\lambda$ being the de Broglie wavelength, so the phase accumulated over a path from ${\bf r}_0$ to ${\bf r}$ is
\begin{equation}
\label{eq:1}
\Phi({\bf r}) = \int_{{\bf r}_0}^{\bf r} \bold{k}\cdot d{\bf r},
\end{equation}
or, since $\lambda = h/p$, where $p$ is particle's momentum, 
\begin{equation}
\label{eq:2}
\Phi({\bf r}) = \frac{1}{\hbar}\int_{{\bf r}_0}^{\bf r} \bold{p}\cdot d{\bf r}.
\end{equation}
This refers to a particular \emph{path}, so the phase difference between neutron beams along two distinct paths is 
\begin{equation}
\Delta\Phi = \frac{1}{\hbar}\int_{{\bf r}_0}^{\bf r} (\bold{p}_{\rm I} - \bold{p}_{\rm II})
\cdot d{\bf r}.
\end{equation}
In our case the path I is the lower route and path II the upper route. The contributions to $\Delta\Phi$ from the vertical parts of these two routes cancel, since the relevant momenta are equal and opposite; and putting $p_{\rm I} = mv$ and $p_{\rm II} = mu$ along the horizontal lower and upper routes respectively (with $v$ and $u$ being the corresponding particle speeds), we find 
\begin{equation}
\Delta\Phi = \frac{1}{\hbar}m(v-u)L,
\end{equation}
where $L$ is the length of the interferometer. 
Conservation of energy now gives us
\begin{equation}
\label{eq:3}
\frac{1}{2} mu^2 = \frac{1}{2} mv^2 - mgH
\end{equation}
where $g$ is the acceleration due to gravity and $H$ the height of the interferometer. Since $gH$ is of the order of $10^{-1}$ ${\rm m}^2{\rm s}^{-2} $ 
and $v^2 \approx 4 \times 10^6$ ${\rm m}^2 {\rm s}^{-2}$ for thermal neutrons, then $gH \ll v^2$, and
\begin{equation}
v-u \approx \frac{gH}{v},
\end{equation}
giving finally
\begin{equation}
\label{eq:5}
\Delta\Phi = \frac{mgA}{\hbar v},
\end{equation}
where $A = LH$ is the area of the interferometer. This phase shift was first predicted and observed in 1975 by Colella, Overhauser and Werner \cite{1}.

    It is pertinent to note that the above expression for the phase shift may
alternatively be obtained by starting from a Lagrangian  $\cal L$  given by
\begin{equation}
\label{eq:6}
{\cal L} = \frac{p^2}{2m} + m{\bf g \cdot r},
\end{equation}
with ${\bf p}$ $( = m{\bf v} = m{\bf {\dot r}})$ defined by 
\begin{equation}
\label{eq:7}
{\bf p} = \frac{\partial {\cal L}}{\partial {\bf \dot{r}}}.
\end{equation}
Equation (\ref{eq:3}) then yields the expected result (\ref{eq:5}).

\section{Simple derivation of the neutron Sagnac effect}
\label{sec:SimpleSAGNAC}

The experiment, first performed by Werner, Staudenmann and Colella 
\cite{2}, measured the effect of the earth's rotation on the neutron 
phase.  To take account of a rotating frame the Lagrangian (\ref{eq:6}) should 
be modified to 
\begin{equation}
\label{eq:8}
{\cal L} = \frac{p^2}{2m} + m{\bf g}\cdot {\bf r} + \bm{\omega}\cdot {\bm{\ell}},
\end{equation}
where $\omega$ is the angular velocity of the frame and ${\bm{\ell}}$ the angular 
momentum of the particle. Then the momentum (\ref{eq:7}) becomes
\begin{equation}
\label{eq:9}
{\bf p} = m {\bf v} + m\bm{\omega} {\bf \times r}. 
\end{equation}
The phase coming from the term in $\omega$ is
\begin{equation}
\label{eq:10}
\Delta \alpha 
= \frac{1}{\hbar} \oint m [\bm{\omega} 
\times {\bf r}]\cdot d{{\bf r}}  = \frac{2m \bm{\omega \cdot {A}}}{\hbar},
\end{equation}
where ${A}$ is again the area of the interferometer. This Sagnac phase is 
typically of the order of $10^{-2}$ of the gravitational COW phase, so 
to detect it means setting up the apparatus in such a way that the COW 
contribution to the phase is zero. This is achieved by having the interferometer 
in a vertical plane --- say in the $r\theta$ or $r\phi$ plane --- and then rotating 
it about a vertical axis. The observations of the phase shift of the neutron due 
to the earth's rotation were found to be in good agreement 
with the theory \cite{2}.



\section{Kerr metric}
\label{sec:Kerr_metric}

We now turn to a general relativistic derivation of the COW and neutron Sagnac effects.
To describe the gravitational field of the rotating earth, we first write down 
the Kerr metric \cite{KERR63} in its standard Boyer-Lindquist form \cite{BL67},
\BWT
\begin{align}
ds^2 &=
\left(1-\frac{r_s}{r}
\frac{1}{1+\left(\frac{a}{r}\right)^2\cos^2\theta}\right)c^2dt^2
+ 2\frac{r_s}{r}\frac{a}{r}\frac{\sin^2\theta }{1+\left(\frac{a}{r}\right)^2\cos^2\theta}
r(cdt) d\varphi
\nonumber \\
& \quad
-\frac{1+\left(\frac{a}{r}\right)^2\cos^2\theta}
{1-\frac{r_s}{r}+\left(\frac{a}{r}\right)^2}dr^2 
-\left(1+\left(\frac{a}{r}\right)^2\cos^2\theta \right)r^2d\theta^2
-\left(
1+\left(\frac{a}{r}\right)^2
+\frac{r_s}{r}\left(\frac{a}{r}\right)^2
\frac{\sin^2\theta }{1+\left(\frac{a}{r}\right)^2\cos^2\theta}
\right)r^2\sin^2\theta d \varphi^2,
\end{align}
\EWT
where $a=(2/5)R^2\w/c$ is the angular momentum parameter,
$r_s = 2GM/c^2$ is the Schwarzschild radius, $M$ is the mass, and $R$ is the radius of the earth. This metric describes the rotating earth as seen from an inertial frame. The
experiments we are considering, however, take place on the earth, and therefore in a
rotating frame, so to find the appropriate metric we must replace  $\varphi$  by  $\varphi'$ 
given by
\BEq
\varphi = \varphi' + \w t ,
\EEq
in which the metric becomes
\BWT
\begin{align}
ds^2 &=
\left\{
1-\frac{r_s}{r}
\frac{1}{1+\left(\frac{a}{r}\right)^2\cos^2\theta}
+ 2\frac{r_s}{r}\frac{a}{r}\frac{r\w}{c}\frac{\sin^2\theta }
   {1+\left(\frac{a}{r}\right)^2\cos^2\theta}
-\frac{r^2\w^2}{c^2}
\left[
1+\left(\frac{a}{r}\right)^2
+\frac{r_s}{r}\left(\frac{a}{r}\right)^2
\frac{\sin^2\theta }{1+\left(\frac{a}{r}\right)^2\cos^2\theta}
\right]\sin^2\theta
\right\}c^2dt^2
\nonumber \\
& \quad
+2\left\{
\frac{r_s}{r}\frac{a}{r}\frac{\sin^2\theta }{1+\left(\frac{a}{r}\right)^2\cos^2\theta}
-\frac{r\w}{c}\left[
1+\left(\frac{a}{r}\right)^2
+\frac{r_s}{r}\left(\frac{a}{r}\right)^2
\frac{\sin^2\theta }{1+\left(\frac{a}{r}\right)^2\cos^2\theta}
\right]\sin^2\theta
\right\}
r(cdt)d\varphi' 
\nonumber \\
& \quad
-\frac{1+\left(\frac{a}{r}\right)^2\cos^2\theta}
{1-\frac{r_s}{r}+\left(\frac{a}{r}\right)^2}dr^2 
-\left[1+\left(\frac{a}{r}\right)^2\cos^2\theta \right]r^2d\theta^2
-\left[
1+\left(\frac{a}{r}\right)^2
+\frac{r_s}{r}\left(\frac{a}{r}\right)^2
\frac{\sin^2\theta }{1+\left(\frac{a}{r}\right)^2\cos^2\theta}
\right]r^2\sin^2\theta d\varphi'^2.
\end{align}
\EWT
This expression is exact. Taking into account that for $r\approx R$ 
(radius of the earth),
$r_s/r\sim 10^{-9}$,
$\w r/{c}\sim 10^{-6}$,
${a}/{r}\sim 10^{-6}$,
${a\w}/{c}\sim 10^{-12}$,
we expand to order $10^{-15}$ and get
\BWT
\begin{align}
\label{eq:expandedMetric00}
ds^2 &=
\left(
1-\frac{r_s}{r}
-\frac{r^2\w^2}{c^2}\sin^2\theta
\right)c^2dt^2
+2\left(
\frac{r_s}{r}\frac{a}{r}
-\frac{r\w}{c}
\right)r\sin^2\theta d\varphi' (cdt) 
\nonumber \\
& \quad
-\left[1+\frac{r_s}{r}-\left(\frac{a}{r}\right)^2\sin^2\theta\right]dr^2 
-\left[1+\left(\frac{a}{r}\right)^2\cos^2\theta \right]r^2d\theta^2
-\left[
1+\left(\frac{a}{r}\right)^2
\right]r^2\sin^2\theta d\varphi'^2.
\end{align}
\EWT

It is convenient to rewrite (\ref{eq:expandedMetric00}) in terms of the ``shifted'' Cartesian coordinates erected on the surface of the earth, by analogy with how it was done in the 
Schwarzschild case in Ref.\ \cite{MOREAU1994}. The idea is to work in a coordinate system whose origin is ``in the laboratory'', on the earth's surface, and also that this should be a Cartesian system, since this simplifies the calculation.
\begin{figure}[!ht]
\centering
\includegraphics[angle=0,width=1.00\linewidth]{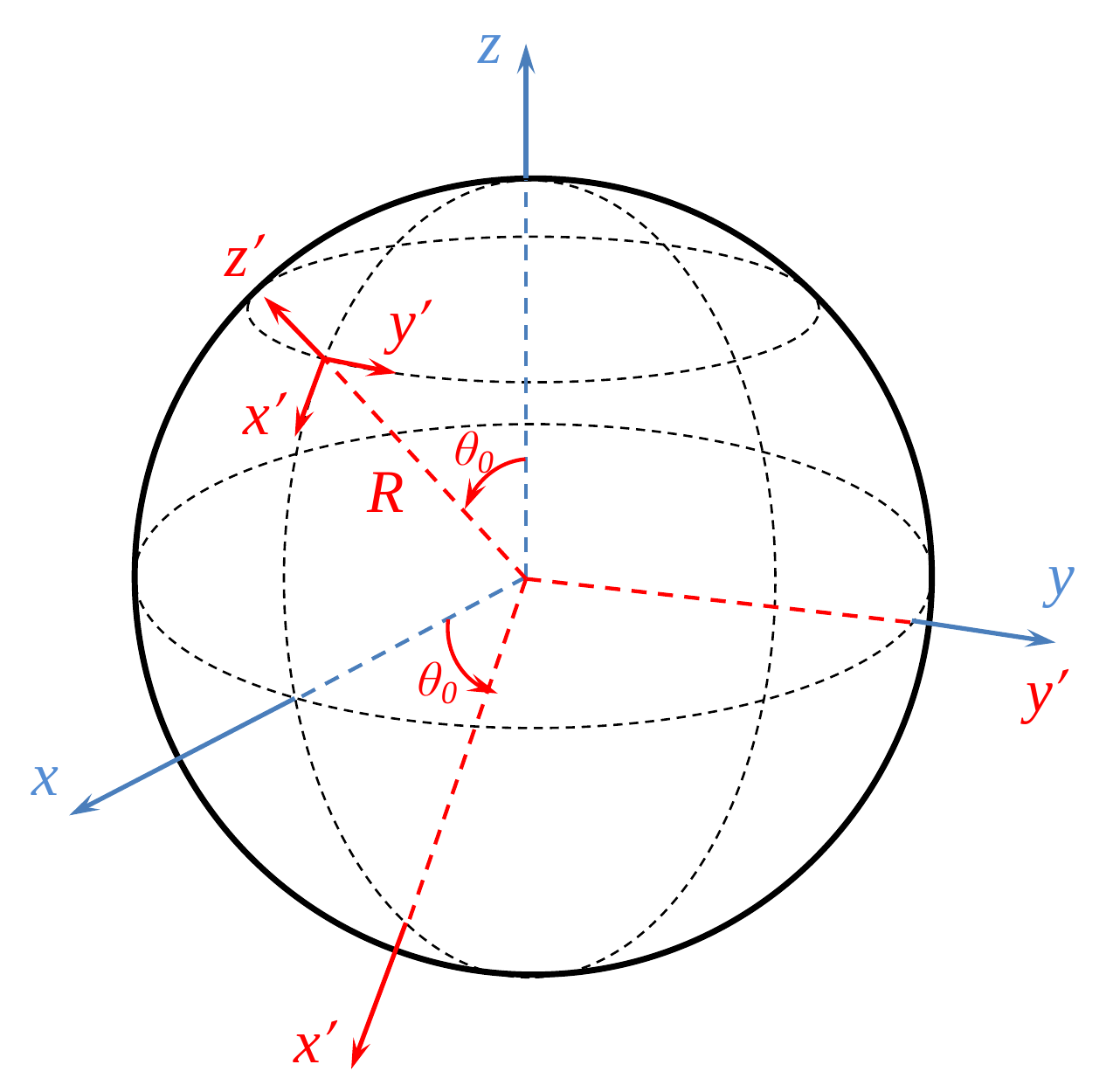}
\caption{ \label{fig:1} 
(Color online.) Local Cartesian coordinates on the surface of the 
rotating Earth.
}
\end{figure}
We first introduce the ``usual'' Cartesian coordinates $(x,y,z)$ defined by
\begin{align}
r = \left(x^2+y^2+z^2\right)^{1/2},
\;
\varphi' =\arctan \frac{y}{x},
\;
\theta =\arccos \frac{z}{r},
\end{align}
and get
\BWT
\begin{align}
\label{eq:expandedMetric01}
ds^2 &=
\left[
1-\frac{r_s}{r}
-\frac{\w^2}{c^2}(x^2+y^2)
\right]c^2dt^2
+2\left(
\frac{r_s}{r}\frac{a}{r}
-\frac{r\w}{c}
\right)  \frac{xdy-ydx}{r} (cdt) 
-\left[
1+\frac{r_s}{r}-\left(\frac{a}{r}\right)^2
\frac{x^2+y^2}{r^2}
\right]
\frac{(xdx+ydy+zdz)^2}{r^2}
\nonumber \\
& \quad
-
\left[
1+\left(\frac{a}{r}\right)^2\frac{z^2}{r^2}
 \right]
\frac{\left(zxdx+zydy-(x^2+y^2)dz\right)^2}{r^2(x^2+y^2)}
-\left[
1+\left(\frac{a}{r}\right)^2
\right]\frac{(xdy-ydx)^2}{x^2+y^2}.
\end{align}
\EWT
We next perform the rotation around the $y$-axis by an angle ${\theta_0}$ (the co-latitude of
interferometer location on the earth's surface; see Fig.\ \ref{fig:1})
and shift the origin by $R$ along the new $z$-axis 
in accordance with
\BEqA
x &=& x'\cos{\theta_0} + (R+z')\sin{\theta_0},
\\
y &=& y',
\\
z &=& -x'\sin{\theta_0} + (R+z')\cos{\theta_0},
\EEqA
and
\BEqA
dx &=& dx'\cos{\theta_0} + dz'\sin{\theta_0},
\\
dy &=& dy',
\\
dz &=& -dx'\sin{\theta_0} + dz'\cos{\theta_0},
\EEqA
where $x'$, $y'$ and $z'$ are the Cartesian coordinates whose origin is on the earth's
surface. We now restrict the experimental region to the neighborhood of this shifted
origin and introduce the weak field approximation, in which
$g_{\m\n} = \eta_{\m\n}+h_{\m\n}$ and $|h_{\mu\nu}| \ll 1$, with  
$\eta_{\m\n}={\rm diag}(1,-1,-1,-1)$. We finally obtain, to
terms linear in $x'/R$, $y'/R$ and $z'/R$,
\begin{align}
\label{eq:h_mn}
h_{00}&= 
-\frac{{r_s} }{R}  \left(1 - \frac{z'}{R}\right)  
\nonumber \\
& \quad
- \frac{{\omega}^2 R^2}{c^2} 
\left[
\left(1 + \frac{2z'}{R}\right)\sin^2\theta_0 + \frac{x'}{R} \sin \left(2\theta_0\right)
\right],
\\
h_{01}&= 
\left(\frac{\omega R}{c} - \frac{a {r_s}}{R^2}\right)\frac{y'}{R}\cos\theta_0,
\\
h_{02}&= 
- \frac{\omega R }{c}
\left[
\left(1  + \frac{z'}{R}\right) \sin\theta_0
+ \frac{x'}{R}\cos\theta_0  
\right]
\nonumber \\
& \quad
+ \frac{a {r_s}}{R^2}
\left[
\left(1 - \frac{2 z'}{R}\right)\sin\theta_0 
+\frac{x'}{R} \cos\theta_0 
\right],
\\
h_{03}&= 
\left(\frac{\omega\, R}{c} - \frac{a {r_s}}{R^2}\right)\frac{y'}{R}\sin\theta_0,
\\
h_{11}&= 
-\frac{a^2}{R^2} 
\left[
\left(1-\frac{2z'}{R} \right)\cos^2\theta_0 - \frac{x'}{R}\sin\left(2 \theta_0\right)
\right],
\\
h_{12}&= 
\frac{1}{2}\frac{a^2}{R^2}\frac{y'}{R}\sin\left(2\theta_0\right),
\\
h_{13}&= 
-\left(\frac{r_s}{R}-\frac{a^2}{R^2}\right)\frac{x'}{R},
\\
h_{22}&= 
- \frac{a^2}{R^2}\left(1 - \frac{2 z'}{R}\right),
\\
h_{23}&= 
- \left(
\frac{{r_s}}{R} -\frac{a^2}{R^2} (1 +  \sin^2\theta_0) 
\right)\frac{y'}{R},
\\
h_{33}&= 
- \frac{{r_s} }{R}\left(1- \frac{z'}{R}\right)
\nonumber \\
& \quad
+\frac{a^2}{R^2} 
\left[
\left(1 - \frac{2z'}{R}\right)\sin^2\theta_0 + \frac{x'}{R} \sin \left(2\theta_0\right)
\right] .
\end{align}


\section{Relativistic derivation of the COW and Sagnac effects}
\label{sec:relativistic_derivation}

We are now in a position to give a relativistic account of the COW and neutron
Sagnac effects. To do so, we need a relativistic expression for the phase shift, which comes from the Feynman-Dirac formula $\exp \left({i}S/{\hbar}\right)$, the amplitude for a particle to travel along a path, with $S = \int \mathcal{L}\,dt$ being the action along the path. The relativistic expression for $S$ is $-mc\int ds$, with $ds^2 = g_{\mu\nu}dx^\mu dx^\nu = c^2 d\tau ^2$, $\tau$ being proper time.  Dividing the expression for $ds^2$ by $ds$ gives
\begin{equation}
ds = g_{\mu\nu} \frac{dx^\mu}{d\tau} dx^\nu = \frac{1}{c} g_{\mu\nu} \frac{dx^\mu}{d\tau} dx^\nu,
\end{equation}
so 
\begin{equation}
S = -m \int g_{\mu\nu} \frac{dx^\mu}{d\tau} dx^\nu = -\int g_{\mu\nu} p^\mu dx^\nu = -\int p_\mu dx^\mu,
\end{equation}
consistent with equation (\ref{eq:2}) above.

We may now proceed, following Stodolsky \cite{5}, by stating 
that the phase $\Phi_{AB}$ accumulated by a particle moving from
spacetime event $A$ to event $B$ is, invoking the weak field approximation,
\begin{align}
\label{eq:phaseAB0}
\Phi_{AB} &= -\frac{mc}{\hbar}
\int_{A}^{B}ds
\nonumber \\
&\approx 
-\frac{mc}{\hbar}
\int_{A}^{B} 
\left(ds_{M}+\frac{1}{2}h_{\m\n}\frac{dx^\m}{ds_{{\cal M}}}dx^\n\right),
\end{align}
where $h_{\m\n}$ is the deviation of the metric $g_{\m\n}$ from its Minkowskian form
$\eta_{\m\n}={\rm diag}(1,-1,-1,-1)$, and 
$ds_{{\cal M}}^2 = \eta_{\r\s}dx^{\r}dx^{\s}$.
Eq.\ (\ref{eq:phaseAB0}) represents the action, normalized to Planck's constant, of a 
{\it freely falling} gravitational probe. 
We assume, as an additional hypothesis, that (\ref{eq:phaseAB0}) can also be applied to 
a probe whose worldline is shaped by, say, a collection of ideally reflecting mirrors that are at rest relative to the chosen coordinate system.
(A mirror is regarded as ideal if on reflection there is no change of particle's energy and 
of the tangential component of its momentum, while the normal component of the
momentum changes sign.) A similar assumption for calculating gravitational effects, 
though in a different context, was made in Ref.\ \cite{GRISHCHUK74}.
The {\it gravitationally induced} phase is then given by
\BEq
\label{eq:phaseAB1}
\Phi_{AB} =  
-\frac{mc}{2\hbar}
\int_{A}^{B}h_{\m\n} u^{\m}_{{\cal M}}dx^{\n},
\EEq
where 
$u^{\m}_{M}=dx^{\m}/ds_{{\cal M}}=(\gamma, \gamma {\bf v}/c)$
is the usual relativistic four-velocity of the particle, ${\bf v}$ is its three-velocity, and $\gamma$ 
is the corresponding gamma-factor. The phase difference between the two interfering paths is then 
\BEq
\label{eq:phaseAB1LOOP}
\Delta \Phi =  
-\frac{mc}{2\hbar}
\oint  h_{\m\n} u^{\m}_{{\cal M}}dx^{\n},
\EEq
where the line integral is taken around the loop formed by the paths.

We now make an important observation that, in the linearized approximation, 
$\sim{\cal O}(h_{\m\n})$, used in Eq.\ (\ref{eq:phaseAB1}), neutron's speed, 
$v\equiv |{\bf v}|$, should be treated as {\it constant}. Any change in the speed  
acquired due to gravity, etc., had already been taken into account when we made the 
linearized approximation (\ref{eq:phaseAB1}). Thus, the gravitationally induced 
phase difference between the interfering paths may be found from the formula
\begin{align}
\label{eq:phaseAB2}
\Delta \Phi 
&=  
-\frac{\gamma mc^2}{2\hbar}
\oint
\biggl[
\left(h_{00}+ \frac{h_{i0}v^i}{c}\right)dt
\nonumber \\
&\quad\quad\quad\quad\quad\quad\quad
+
\left(h_{0j}+\frac{h_{ij} v^i}{c}\right)\frac{dx^{j}}{c}
\biggr],
\end{align}
where $v=\sqrt{\d_{ij}v^iv^j}$ is regarded as constant.

Eq.\ (\ref{eq:phaseAB2}) represents the accumulated phase difference for a {\it single} 
orientation of the loop. This phase difference, which we call intrinsic, is not directly 
observable. In an actual experiment, at least {\it two} orientations are involved, 
and it is the {\it shift} in the intrinsic phase difference during the rotation of the 
loop from one position to the other that is experimentally measurable.

Assuming that the loop is a {\it rectangle} placed in the $x'y'$-plane, with the sides 
parallel to the $x'$ and $y'$ axes, we have $z'=0$ and ${\bf v} = (v_x,v_y,0)$, and 
upon using (\ref{eq:phaseAB2}), find the corresponding intrinsic phase difference,
\BWT
\begin{align}
\label{eq:x'y'-plane-phase}
(\Delta \Phi)_{x'y'} 
&=  
-\frac{\gamma mc^2}{2\hbar}
\left\{
\int_{(0,0,0)}^{(\Delta x', 0,0)}
-
\int_{(0, \Delta y',0)}^{(\Delta x', \Delta y',0)}
\right\}
\left[
\left(h_{00}+ \frac{h_{10}v}{c}\right)dt
+
\left(h_{01}+\frac{h_{11} v}{c}\right)\frac{dx'}{c}
\right]
\nonumber \\
&\quad  
-\frac{\gamma mc^2}{2\hbar}
\left\{
\int_{(\Delta x',0,0)}^{(\Delta x', \Delta y',0)}
-
\int_{(0,0,0)}^{(0, \Delta y',0)}
\right\}
\left[
\left(h_{00}+ \frac{h_{20}v}{c}\right)dt
+
\left(h_{02}+\frac{h_{22} v}{c}\right)\frac{dy'}{c}
\right]
\nonumber \\
&=  
-\frac{\gamma mc^2}{2\hbar}
\left\{
\int_{(0,0,0)}^{(\Delta x', 0,0)}
-\int_{(0, \Delta y',0)}^{(\Delta x', \Delta y',0)}
\right\}
\left(
\frac{h_{00}}{v}+ \frac{2h_{10}}{c}+\frac{h_{11} v}{c^2}
\right)dx'
\nonumber \\
&\quad  
-\frac{\gamma mc^2}{2\hbar}
\left\{
\int_{(\Delta x,0,0)}^{(\Delta x', \Delta y',0)}
-\int_{(0,0,0)}^{(0, \Delta y,0)}
\right\}
\left(
\frac{h_{00}}{v}+ \frac{2h_{20}}{c}+\frac{h_{22} v}{c^2}
\right)dy'
\nonumber \\
&=    
+\frac{\gamma mc^2}{2\hbar v}
\left\{
 \frac{4v}{c}\left(\frac{\omega R}{c} - \frac{a {r_s}}{R^2}\right)\cos\theta_0
+
\frac{{\omega}^2 R^2}{c^2} 
 \sin \left(2\theta_0\right) 
\right\}
\frac{\Delta x' \Delta y'}{R},
\end{align}
\EWT
which vanishes in the $a, \omega \rightarrow 0$ limit, as had to be expected. 
In a similar manner, for the $z'x'$ and $y'z'$ orientations,
we get
\BWT
\begin{align}
\label{eq:z'x'-plane-phase}
(\Delta \Phi)_{z'x'} 
&=  
\frac{\gamma mc^2}{2\hbar v}
\biggl[
\frac{{r_s} }{R}
- 
\frac{{\omega}^2 R^2}{c^2} 
\left(2\sin^2\theta_0 - \sin \left(2\theta_0\right)\right)
+
\frac{v^2}{c^2}
\frac{a^2}{R^2} 
\left(2\cos^2\theta_0  -\sin \left(2\theta_0\right)\right)
\biggr]\frac{\Delta z' \Delta x'}{R},
\end{align} 
\begin{align}
\label{eq:y'z'-plane-phase}
(\Delta \Phi)_{y'z'} 
&=  
\frac{\gamma mc^2}{2\hbar v}
\biggl[
\frac{{r_s} }{R} - \frac{2{\omega}^2 R^2}{c^2} \sin^2\theta_0
- 
\frac{2v}{c}
\left(
\frac{2\omega R }{c}+ \frac{a {r_s}}{R^2}
\right)\sin\theta_0 
+
\frac{v^2}{c^2}
 \frac{2a^2}{R^2}
\biggr]\frac{\Delta y'\Delta z'}{R}.
\end{align} 
\EWT
Combining Eqs.\ (\ref{eq:x'y'-plane-phase}) and (\ref{eq:z'x'-plane-phase}),
and assuming that the loop is now rotated around the $x'$-axis from horizontal $x'y'$
 to vertical $z'x'$ position, we get, using $\Delta x' \equiv L$ and 
$\Delta y' =\Delta z' \equiv H$, the experimentally observable
COW change of phase,
\BWT
\begin{align}
\label{eq:rtheta-plane-phase}
(\Delta \Phi)_{{\rm COW}} &\equiv (\Delta \Phi)_{z'x'} - (\Delta \Phi)_{x'y'}
\nonumber \\
&=
\frac{\gamma mc^2}{2\hbar v}\frac{LH}{R}
\biggl\{
\frac{{r_s} }{R}
-
\frac{4v}{c}\left(\frac{\omega R}{c} - \frac{a {r_s}}{R^2}\right)\cos\theta_0
- 
\frac{2{\omega}^2 R^2}{c^2} \sin^2\theta_0 
+
\frac{v^2}{c^2}
\frac{a^2}{R^2} \left(2\cos^2\theta_0 -\sin \left(2\theta_0\right)\right)
\biggr\}
\nonumber \\
&= \gamma \frac{mgA}{\hbar v} + ({\rm terms\; in\;} \omega {\rm \; and\;} \omega^2),
\end{align}
\EWT
where we have made the identification
\BEq
\frac{r_s}{R^2}\equiv \frac{2g}{c^2},
\EEq
with $g$ being the acceleration due to gravity at the earth's surface. It therefore
turns out that in this relativistic formulation the COW phase shift is merely the
simple result (\ref{eq:5}), corrected by the factor $\gamma$, and beyond that, further
corrected (slightly surprisingly!) by terms resulting from the rotation of the
earth. These terms are two or more orders of magnitude smaller than the first terms
in (\ref{eq:rtheta-plane-phase}): 
$r_s / R = 1.4 \times 10^{-9}$,  $4v\omega R/c^2 = 4.7 \times 10^{-11}$, 
$\omega R/c = 1.6 \times 10^{-6}$, $ar_s/R^2 = 0.9 \times 10^{-15}$, 
where we have taken $v = 2.2 \times 10^{3}$ m/s for thermal neutrons.

On the other hand, combining Eqs.\ (\ref{eq:z'x'-plane-phase}) 
and (\ref{eq:y'z'-plane-phase}),
and assuming that the loop is rotated around the $z'$-axis 
from vertical $z'x'$ to vertical $y'z'$
position, we get, using $\Delta x' = \Delta y' \equiv L$ 
and $\Delta z' \equiv H$, the phase shift
\BWT
\begin{align}
\label{eq:Sagnac-phase}
(\Delta \Phi)_{\rm Sagnac} &\equiv (\Delta \a)_{y'z'} - (\Delta \a)_{z'x'}
\nonumber \\
&=
+\frac{\gamma mc^2}{2\hbar v}
\frac{LH}{R}
\biggl\{
- 
\frac{2v}{c}
\left(
\frac{2\omega R }{c}+ \frac{a {r_s}}{R^2}
\right)\sin\theta_0 
-
\frac{{\omega}^2 R^2}{c^2}  \sin \left(2\theta_0\right)
+
\frac{v^2}{c^2}
\frac{a^2}{R^2} 
\left(2\sin^2\theta_0  + \sin \left(2\theta_0\right)\right)
\biggr\}
\nonumber \\
&= - \gamma \frac{2m\omega A}{\hbar} + ({\rm terms\; in\;} a, a^2, \omega^2).
\end{align}
\EWT
We see, similarly to the COW case, that the magnitude of the Sagnac effect is the
same as obtained in the simple derivation, corrected by $\gamma$, and modified by
considerably smaller terms.


\section{Summary}

We conclude that by making the weak field approximation we may straightforwardly derive expressions for the COW and neutron Sagnac phase shifts. Our general relativistic calculation yields the same results as simple semi-classical arguments do, corrected only by the relativistic factor $\gamma$, and by higher order terms involving the angular velocity of the earth.


\end{document}